# Storing light as a mechanical excitation in a silica optomechanical resonator

Victor Fiore[1], Yong Yang[1], Mark Kuzyk[1], Russell Barbour[1], Lin Tian[2], and Hailin Wang[1]

[1]Department of Physics, University of Oregon, Eugene, Oregon 97403, USA
[2]Department of Physics, University of California, Merced, California 95343, USA

Abstract

We report the experimental demonstration of optomechanical light storage in a silica resonator. We use writing and readout laser pulses tuned to one mechanical frequency below an optical cavity resonance to control the coupling between the mechanical displacement and the optical field at the cavity resonance. The writing pulse maps a signal pulse at the cavity resonance to a mechanical excitation. The readout pulse later converts the mechanical excitation back to an optical pulse. The light storage lifetime is determined by the relatively long damping time of the mechanical excitation.

Light is a natural and ideal information carrier, but is difficult to store. Light storage is important for all-optical information networks and is also an essential ingredient for long-distance quantum communication[1,2]. A variety of approaches for light storage have been actively pursued. Storage of light as spin excitations in atomic media or as persistent atomic excitations in inhomogeneously broadened solids has been realized in quantum as well as classical regimes[3-6]. Optical pulses have also been stored in dynamically-tunable coupled-resonator optical waveguides or as acoustic excitations in optical fibers[7-10].

Optomechanical resonators, in which optical fields couple to mechanical oscillations via radiation pressure (see Fig. 1a), provide another potential avenue for light storage. Optomechanical interactions have been successfully explored for the control of mechanical as well as optical processes in these resonators in the steady state[11]. Earlier experimental studies have demonstrated optomechanical parametric amplification, laser cooling, and normal mode splitting of a mechanical mode[12-18]. Optomechanical processes analogous to the well known phenomenon of electromagnetically-induced transparency (EIT) have also been realized recently in both optical and microwave regimes[19-21].

Here, we report a proof-of-principle experimental demonstration of storing light as a mechanical excitation in a silica resonator via transient optomechanical processes, with the storage lifetime determined by the relatively long decay time of the mechanical excitation. In comparison with atomic or spin systems, an optomechanical resonator features the remarkable property that an optically-active mechanical mode can couple to any of the optical resonances via radiation pressure. Optomechanical processes not only can store light at a given wavelength as a mechanical excitation, but also can map the stored mechanical excitation back to light at practically any desired wavelength[22-25]. This capability of wavelength conversion can play a special role in both classical and quantum networks, for example, by converting optical information from a given wavelength, including microwaves, to a wavelength that is suitable for long distance communication, or by mapping photons emitted from one type of quantum system to photons that can couple to another type of quantum system.

In addition, an optical pulse can also be localized and its spatial-temporal profile be stored in an array of optomechanical resonators via processes analogous to those used in dynamically tunable coupled-resonator optical waveguides[7], as proposed recently[26]. In comparison with an all-optical system, optical properties of an optomechanical system can be effectively controlled with optical pulses via radiation pressure. Mechanical oscillators with high quality factors can also feature a storage lifetime much longer than that can be achieved with optical micro- or nano-resonators.

For optomechanical light storage, we use "writing" and "readout" laser pulses tuned to one mechanical frequency, $\omega_m$, below the optical cavity resonance to control the coupling between the mechanical displacement and the optical field at the cavity resonance (see Fig. 1b). The writing pulse maps a signal pulse at the cavity resonance to a mechanical excitation. The readout pulse later converts the mechanical excitation back to an optical pulse at the cavity resonance. As illustrated in Fig. 1a, in an optomechanical resonator, the displacement of a mechanical oscillator modulates the frequency of an optical cavity mode, with $\omega_c(x) = \omega_0 + gx$, where $x$ is the mechanical displacement and $g$ is the optomechanical coupling coefficient. In the limit that $\omega_m \gg \kappa$ (the resolved sideband limit) and $n_c \gg 1$ where $\kappa$ is the cavity decay rate and $n_c$ is the intracavity photon number for either the writing or the readout pulse, the interaction between the mechanical displacement and the optical field at the cavity resonance can be approximated as a coupled oscillator system, with an effective interaction Hamiltonian given by $H_{\text{int}} = \hbar G(\hat{a}^+\hat{b} + \hat{a}\hat{b}^+)$, where $\hat{a}$ and $\hat{b}$ are the annihilation operators for the optical field and the mechanical displacement, respectively, and $G = gx_{zpf}\sqrt{n_c}$ is the effective optomechanical coupling rate with $x_{zpf}$ being the zero-point fluctuation for the mechanical mode[22]. In this system, the writing and readout pulses can control or switch on/off the effective optomechanical coupling between the optical field and the mechanical displacement. As shown in Fig. 1c, for the storage process, a writing pulse couples a signal pulse to the mechanical mode, generating a mechanical excitation. For the retrieval process, a readout pulse couples the stored mechanical excitation to the cavity mode, mapping the mechanical excitation back to an optical pulse.

We used silica microspheres as a model system for an optomechanical resonator[18]. Deformed silica microspheres with a deformation near 2% and a diameter near 30 μm were fabricated by fusing together two non-deformed microspheres of similar sizes with a $CO_2$ laser. The small deformation enables free-space evanescent excitation of whispering gallery modes (WGMs) near the sphere equator, with a coupling efficiency of 9% for resonant excitation. A breathing mode of the sphere with (*n, l*) = (1, 2), where *n* and *l* are the radial and angular mode numbers, respectively, was used as the mechanical oscillator. For the experiments presented in this paper, $\omega_m/2\pi$ = 108.4 MHz, $\gamma_m/2\pi$ = 38 kHz, and $\kappa/2\pi$ = 40 MHz, as determined from displacement power spectra and optical transmission spectra.

The writing, readout, and signal pulses with a wavelength near 800 nm were all derived from the same laser beam generated from a tunable Ti:Sapphire ring laser. The writing and readout pulses, with the same frequency, $\omega_l$, were obtained by gating the laser beam with an acousto-optic modulator (AOM). The laser pulses propagated through an electro-optic modulator (EOM), with the electro-optic phase modulation synchronized with the writing pulse (there is no phase modulation for the readout pulse). The higher frequency sideband generated by the phase modulation served as the signal pulse. The frequency of the signal pulse, $\omega_s$, is locked to a given WGM resonance with a Pound-Drever-Hall technique. The timing and temporal profile of the signal pulse is the same as those of the writing pulse. The intracavity peak power of the signal pulse is kept below 1% of that of the writing and readout pulses. Unless otherwise specified, we set the EOM modulation frequency to $\omega_m$ such that $\omega_l = \omega_s - \omega_m$ and have used the pulse sequence shown in the inset of Fig. 2a. All experiments were carried out at room temperature.

Incident writing and readout pulses were also used as the local oscillator for the heterodyne detection of signal and retrieved pulses emitted from the optical resonator, respectively. Since the signal pulse is generated directly from the writing pulse with an electro-optic phase modulation, the heterodyne detection is not sensitive to the part of the signal pulse that is not emitted from the optical resonator. A spectrum analyzer in a time-gated detection mode was used for time-resolved heterodyne detection, with the time resolution limited by the resolution bandwidth (1 MHz) as well as the gate length (3 µs). More experimental details along with a diagram for the experimental setup are described in the supplementary materials.

Figure 2 presents a proof-of-principle experimental demonstration of optomechanical light storage. Figure 2a shows the heterodyne-detected signal and retrieved pulses emitted from the silica resonator. In this experiment, the readout pulse, which arrives 6.5 µs after the center of the signal pulse, interacts with the mechanical excitation induced by the signal and the writing pulse, generating the retrieved pulse. The incident writing and readout pulses produce an estimated intracavity photon number of $n_c = 1.5 \times 10^6$, corresponding to a peak optomechanical coupling rate of $G_0/2\pi = 2$ MHz. To determine the light storage lifetime, we plot in Fig. 2b the energy of the retrieved pulse, obtained from measurements similar to those shown in Fig. 2a, as a function of the delay between the writing and the readout pulse. The pulse energy decays exponentially as a function of the delay, yielding a storage lifetime of 3.5 µs, which is in good agreement with the mechanical linewidth, $\gamma_m/2\pi = 38$ kHz, obtained from the displacement power spectrum.

The temporal profiles of the signal and retrieved pulses shown in Fig. 2a are significantly modified by the time resolution of the heterodyne detection measurement and by the temporal profile of the local oscillators (i.e., the writing and readout pulses). Figure 2c shows the heterodyne-detected retrieved pulse with the durations of the readout pulse increasing incrementally from 0.3 µs to 1.4 µs, indicating that the time-resolution of the heterodyne measurement is approximately 3 µs. The powers obtained in these transient measurements are effectively the average power over a given detection period.

Optomechanical storage and retrieval processes are characterized by their distinct dependence on the intensity of the writing and readout pulses and on the detuning between the signal and writing/readout pulses. Figure 3a shows the dependence of the retrieved pulse energy on the relative readout intensity, $I/I_0$, with $I_0$ corresponding to $G_0/2\pi=0.7$ MHz. Similar highly nonlinear dependence was also observed when the writing intensity was varied. Figure 3b shows the retrieved pulse energy as a function of the detuning between the signal and the writing/readout pulses at two different readout intensities, with $\omega_s$ fixed at the cavity resonance. For Figs. 3a and 3b, the writing intensity is fixed at $I_0$ and the duration of the readout pulse is 3 µs. The observed resonance in Fig. 3b centered at $\omega_s - \omega_l = \omega_m$ confirms the optomechanical origin of the light storage and retrieval processes. As shown in Fig. 3b, the spectral lineshape observed is independent of the intensity of the readout pulse. The same spectral lineshape was also observed at higher readout intensities.

The experimental results in Figs. 2 and 3 are in good agreement with the theoretical calculation, for which we used the coupled oscillator equations to describe the optomechanical coupling between the mechanical displacement and the optical field at the cavity resonance (see the supplementary material), with all parameters determined by the experiments. Figure 3c shows the calculated intracavity intensity for both the signal and the retrieved pulse, along with the intensity of the stored mechanical oscillation, as a function of time under the conditions of the experiment shown in Fig. 2a. Although the temporal profiles of the pulses are significantly modified in the experiment due to the limited time-resolution as well as the temporal profile of the local oscillators, the relative pulse energy obtained in the experiment agrees well with the theoretical expectation. Figure 3c also indicates that the small signal-to-retrieval conversion efficiency observed is to a large degree due to the decay of the mechanical excitation, since the delay between the readout and writing pulses is nearly twice the mechanical lifetime.

The signal-to-retrieval conversion efficiency shown in Fig. 2a is also limited by the relatively large cavity decay rate of the deformed silica resonator. Near unity conversion efficiency can be realized with $G \gg (\gamma_m, \kappa)$, which is achievable with non-deformed silica resonators that feature ultrahigh optical finesse. For the coupled oscillator system, the mapping between an optical pulse and a mechanical excitation is governed by the optomechanical pulse area defined as $\theta(t) = \int dt G(t)$ [22]. In the absence of any damping, perfect mapping occurs with $\pi/2$ writing and readout pulses. The theoretical calculation shown in Fig. 1c, which is obtained with $\pi/2$ writing and readout pulses and with $\kappa/2\pi = 0.5$ MHz, $\gamma_m/2\pi = 10$ kHz, and $G_0/2\pi = 2$ MHz, features a much greater conversion efficiency than that in Fig. 3c. For our experiments with $\kappa \gg G \gg \gamma_m$, the conversion efficiency continues to rise as the pulse areas exceed $\pi/2$, but eventually saturates, leading to the observed nonlinear dependence on the intensities of the writing and readout pulses. As shown in Fig. 3a, the observed dependence of the retrieved pulse energy on the readout intensity is well described by the theoretical calculation. Similar behaviors have also been obtained for the dependence on the intensity of the writing pulse.

The optomechanical storage process is closely related to EIT, but with the coherent coupling between the signal field and the mechanical displacement controlled by the optomechanical area of the writing pulse. In comparison, the optomechanical retrieval process is closely related to optomechanical laser cooling. For the retrieval process, the readout laser pulse couples to the stored mechanical excitation, transferring or damping the mechanical excitation to generate the retrieved pulse, as shown in Figs. 1c and 3c.

The optomechanical retrieval process features behaviors that are qualitatively different from those of EIT. Figure 3b plots as the dotted curve the theoretically-calculated retrieved pulse energy as a function of $\omega_s - \omega_l - \omega_m$. The theoretical spectral lineshape is in good agreement with the experiment. Note that the calculated lineshape and the corresponding spectral linewidth remain independent of the readout intensity even at high readout intensities such that $4G^2/\kappa$, which is the readout-induced mechanical damping rate, far exceeds the spectral linewidth (see supplementary materials). In comparison, for an EIT process, a strong control beam broadens the linewidth of the transparency window by $4G^2/\kappa$ [19, 21]. Additional calculations show that the observed spectral linewidth in Fig. 3b is primarily due to the finite duration of the signal and writing pulses.

In summary, by exploiting transient optomechanical processes, we have successfully demonstrated optomechanical light storage in a silica resonator, with the storage lifetime determined by the mechanical damping time. At room temperature, thermal excitation of the mechanical oscillator leads to a thermal background in the storage and retrieval processes, which is negligible for most classical applications, but prevents the application of the optomechanical light storage in a quantum regime. Recent experimental efforts have successfully cooled mechanical oscillators to its quantum ground state[27]. Together with these latest advances, the demonstration of optomechanical light storage opens up exciting opportunities for exploring unique properties of optomechanical systems in applications including quantum memory and quantum optical wavelength conversion.

This work is supported by the DARPA-MTO ORCHID program through a grant from AFOSR.

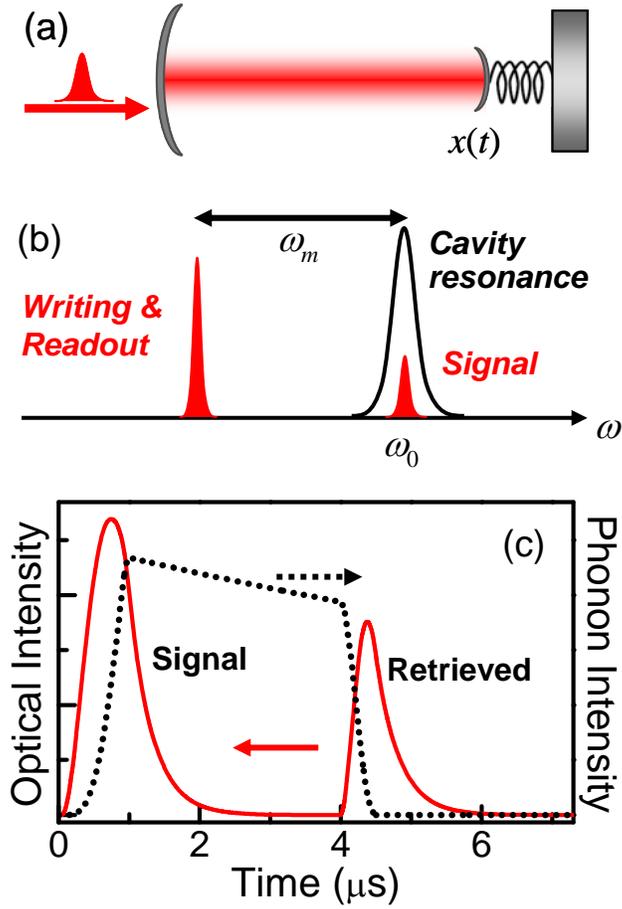

FIG 1. (Color online) (a) Schematic of an optomechanical resonator. (b) Spectral position for the writing, readout, and signal pulses. (c) The intensity of intracavity signal and retrieved pulses (solid red), along with the intensity of the stored mechanical oscillation (dotted black), as a function of time, illustrating the optomechanical process of light storage and retrieval.

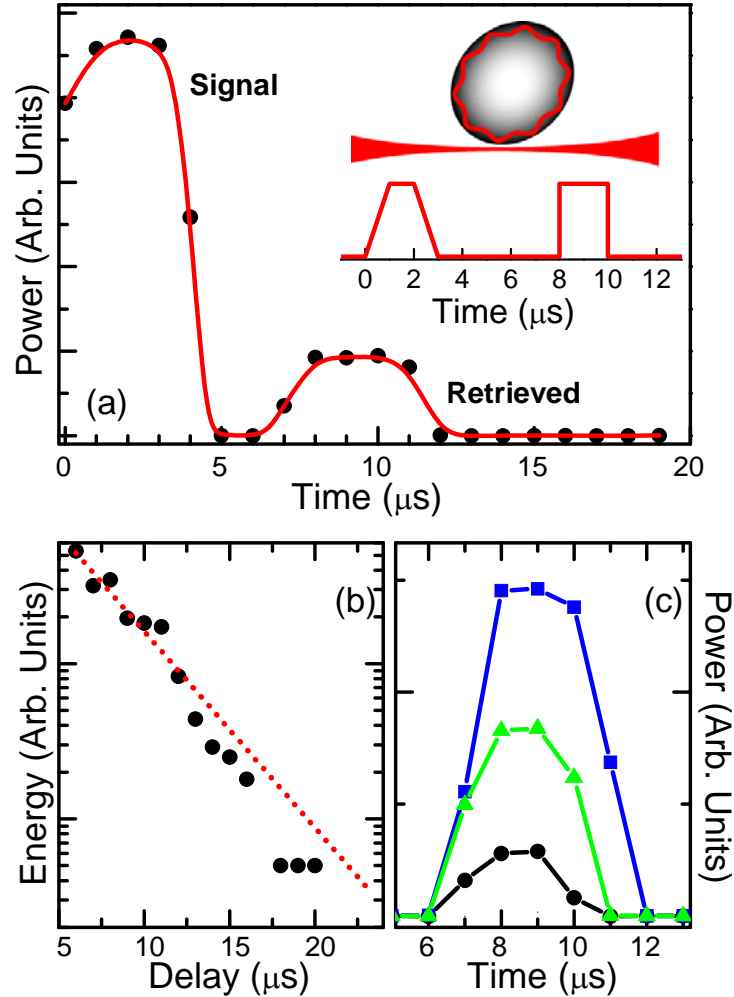

FIG 2. (color online) (a) The power of the heterodyne-detected signal and retrieved pulses emitted from the resonator as a function of time. The solid curve serves as a guide to the eye. The inset illustrates free-space evanescent excitation of a WGM and the timing of the writing and readout pulses used for the experiment. (b) Retrieved pulse energy as a function of the delay between the readout (3 μs in duration) and writing pulses. An exponential fit shown as the red dotted line yields a storage lifetime of 3.5 μs. (c) The temporal profile of the retrieved pulse generated by a readout pulse with duration of 0.3, 0.6, 1.4 μs (from bottom to top), under otherwise identical condition as those for (a).

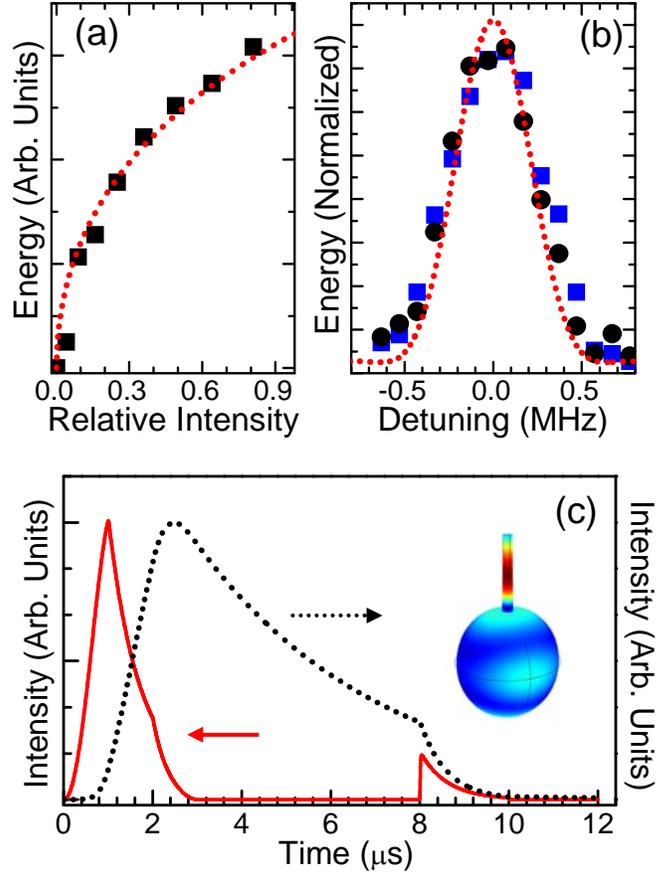

FIG 3. (color online) (a) Retrieved pulse Energy as a function of the relative readout intensity. (b) Retrieved pulse energy (normalized to that at zero detuning) as a function of $\omega_s - \omega_l - \omega_m$ and with a relative readout intensity of 1 (blue square) and 0.45 (black circle). The red dotted lines in (a) and (b) show the theoretical calculations with all parameters taken from the experiment. (c) Theoretically calculated temporal profile of the signal and retrieved pulses, along with the intensity of the stored mechanical oscillation (dotted line), with all parameters taken from the experiment in Fig. 2a. The inset shows the calculated spatial displacement pattern of the (1, 2) mechanical mode.

[25] C. A. Regal and K. W. Lehnert, Journal of Physics: Conference Series **264**, 012025 (2011).

[26] D. E. Chang *et al.*, New Journal of Physics **13**, 26 (2011).

[27] A. D. O'Connell *et al.*, Nature **464**, 697 (2010).- 12 -

# Supplementary Materials

## 1. Theoretical model

We consider the optomechanical coupling between a mechanical oscillator and a single optical cavity mode with an interaction Hamiltonian given by[1]

$$H_{int} = \hbar g \hat{a}^+ \hat{a} \hat{x} \tag{1}$$

where $\hat{x} = x_{zpf}(\hat{b} + \hat{b}^+)$, $\hat{a}$ and $\hat{b}$ are the annihilation operator for the optical cavity mode and the mechanical mode, respectively, $x_{zpf} = \sqrt{\hbar/2m\omega_m}$ with $m$ and $\omega_m$ being the effective mass and frequency of the mechanical oscillator, respectively, is the zero point fluctuation for the mechanical mode, $g = \partial \omega_c / \partial x$ with $\omega_c$ being the cavity resonance frequency. We assume that the intracavity optical field consists of a weak signal field with frequency $\omega_s$ near the cavity resonance and a control field with frequency $\omega_l$ either blue- or red-detuned from the cavity resonance by an amount near $\omega_m$.

For a strong control field, we make the mean-field approximation, in which we treat the control field classically and linearize the optomechanical coupling with respect to the signal field. Effectively, we have

$$\hat{a} \to \sqrt{n_c} e^{-i\omega_l t} + \hat{a}_s.$$

where $n_c$ is the intracavity photon number for the control beam and $\hat{a}_s$ is the annihilation operator for the signal field. The linearized Hamiltonian is given by

$$H_{int} \approx \hbar G(e^{-i\omega_l t}\hat{a}_s^+ \hat{b} + e^{i\omega_l t}\hat{a}_s \hat{b}^+) + \hbar G(e^{-i\omega_l t}\hat{a}_s^+ \hat{b}^+ + e^{i\omega_l t}\hat{a}_s \hat{b}) \tag{2}$$

where $G = gx_{zpf}\sqrt{n_c}$ is the effective optomechanical coupling rate and $\hat{b}$ should now be viewed as the annihilation operator for the renormalized phonons. The first term in Eq. 2 has the form of the well-known beam-splitter Hamiltonian in quantum optics and the second term has the form of the well-known parametric down conversion Hamiltonian. As shown in Eq. 2, the strong control field mediates and controls the optomechanical coupling between the signal field and the mechanical displacement.

The beam-splitter Hamiltonian can be used for optomechanical quantum state transfer[2-5]. In the resolved-sideband limit and with the control field tuned to $\omega_m$ below the cavity resonance, the first term in Eq. 2 dominates the optomechanical coupling. We thus consider only the beam-splitter Hamiltonian. In the absence of any damping processes, the signal field and the mechanical oscillator evolve according to

$$\hat{a}_s(t) = [\hat{a}_s(0)\cos\theta(t) - i\hat{b}(0)\sin\theta(t)]e^{-i\omega_s t}$$
$$\hat{b}(t) = [\hat{b}(0)\cos\theta(t) - i\hat{a}_s(0)\sin\theta(t)]e^{-i\omega_m t}$$
(3)

where $\theta(t) = \int dt G(t)$ is the optomechanical pulse area and we also assume $\omega_s - \omega_l = \omega_m$. After an optomechanical "$\pi/2$ pulse" with $\theta(\tau) = \pi/2$, we have $\hat{a}_s(\tau) = -i\hat{b}(0)e^{-i\omega_s \tau}$ and $\hat{b}(\tau) = -i\hat{a}_s(0)e^{-i\omega_m \tau}$, which maps the motional state to the optical state and the optical state to the motional state. In this regard, the "$\pi/2$ pulses" perform the essential function of the optomechanical storage and retrieval[5].

For simplicity, the effective interaction Hamiltonian in the main text is written with $\hat{a}$ and $\hat{b}$ as the annihilation operators for the signal field and the mechanical displacement in their respective rotating frames.

With optical and mechanical damping processes included, the optomechanical coupling between the signal field and mechanical displacement can be described by the following equations of motion (which are Langevin equations for the expectation values and are of the form of the classical coupled oscillator equations):

$$\dot{\alpha} = -[i(\omega_0 - \omega_s) + \kappa/2]\alpha - iG\beta + \sqrt{\kappa_{ex} P_{in}}$$
$$\dot{\beta} = -[i(\omega_m + \omega_l - \omega_s) + \gamma_m/2]\beta - iG\alpha$$
(4)

where $\alpha = <\hat{a}_s> \exp(i\omega_s t)$ and $\beta = <\hat{b}> \exp[i(\omega_s - \omega_l)t]$ determines the amplitudes of the intracavity signal field and the mechanical displacement, respectively, $\omega_0$ is the resonance frequency of the un-displaced cavity mode, $P_{in}$ is the photon flux of the incident signal field, and $\kappa_{ex}$ is the relevant output coupling rate of the cavity.

In the absence of damping, the mapping between the signal field and the mechanical displacement follows Eq. 3. Perfect mapping occurs with $\pi/2$ writing and readout pulses, as discussed above. In the bad cavity limit with $\kappa \gg G \gg \gamma_m$, the signal-to-retrieval conversion efficiency continues to rise even when the pulse areas far exceed $\pi/2$. Figure 1 shows the calculated retrieved pulse energy as a function of the optomechanical area as well as the intensity of the readout pulse. Note that the saturation of the conversion efficiency at high readout intensities is more apparent in the intensity dependence.

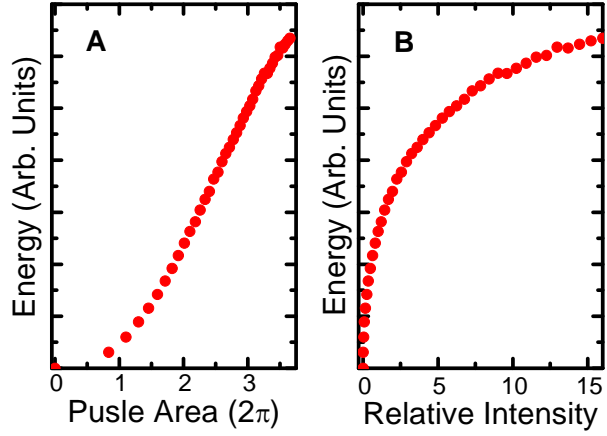

FIG. 1  Retrieved pulse energy as a function of the optomechanical area (A) and relative intensity (B) of the readout pulse. Relative intensity of 1 corresponds to a peak optomechanical coupling rate of $G_0/2\pi = 0.7$ MHz. The parameters used are same as those for Fig. 3a in the main manuscript.

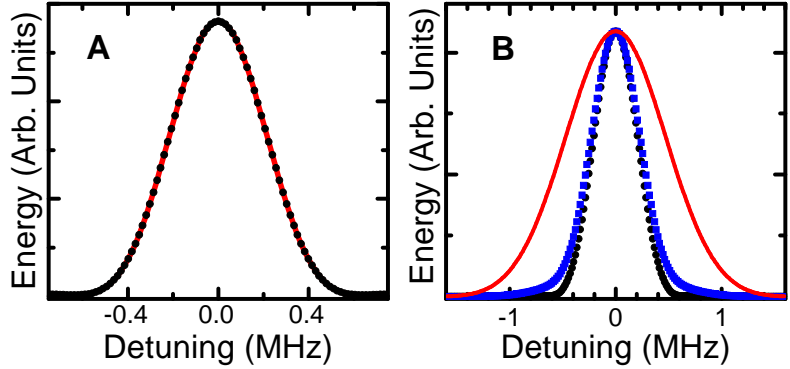

FIG. 2  Retrieved pulse energy (normalized to that at zero detuning) as a function of detuning, $\omega_s - \omega_l - \omega_m$. (a) The readout intensity corresponds to $G_0/2\pi = 0.7$ MHz (black dots) and 5 MHz (red line), with the writing intensity corresponding to $G_0/2\pi = 0.7$ MHz. (b) The writing intensity corresponds to $G_0/2\pi = 0.7$ MHz (black dots), 2 MHz (blue square), and 5 MHz (red line). Other parameters are the same as those in Fig. 3b of the main manuscript.

Figure 2 shows the calculated retrieved pulse energy (normalized to that at zero detuning) as a function of detuning, $\omega_s - \omega_l - \omega_m$. The spectral lineshape is independent of the intensity of the readout pulse. As an example, Fig. 2a shows the spectral lineshape obtained with $G_0/2\pi =$ 0.7 MHz and $G_0/2\pi = 5$ MHz. At the higher intensity, the readout-induced mechanical damping rate $4G_0^2/\kappa = 1.25$ MHz, considerably exceeds the linewidth (~ 0.5 MHz) of the spectral lineshape. In contrast, Fig. 2b shows the broadening of the lineshape by a strong writing pulse. Note that for the transient response, the broadening observed is somewhat smaller than $4G_0^2/\kappa$, expected for the steady-state response of electromagnetically-induced transparency (EIT)[6, 7]. These results also indicate the measured spectral linewidth in Fig. 3b of the main manuscript is primarily due to the finite duration of the signal and writing/readout pulses.

## 2. Fabrication of silica microspheres

Deformed silica microspheres were fabricated by fusing together two regular microspheres of similar sizes with a focused $CO_2$ laser beam. The degree of deformation was controlled through repeated heating. For free space excitation of whispering gallery modes (WGMs), the silica microsphere was held in a vertical position attached to a thin stem. The excitation laser beam was in the horizontal equatorial plane and was focused onto a region, which is 45 degrees away from a symmetry axis. More details of the fabrication and free space evanescent excitation of WGMs can be found in Ref. 8 and its supplemental materials.

## 3. Experimental setup

The detailed setup for the optomechanical light storage experiment is illustrated in Fig. 3. The three incident laser pulses, including the writing, readout, and signal were all derived from a single-frequency CW Ti:Sapphire ring laser (Coherent 899-21). The laser beam was first sent through an acousto-optic modulator (AOM). The first order beam diffracted from the AOM then propagated through an electro-optic modulator (EOM). An arbitrary waveform generator was used to gate the AOM and generate the writing and readout pulse sequence. Another arbitrary waveform generator was used to gate the EOM. The EOM operated at or near the frequency of the relevant mechanical mode of the silica microsphere. The higher frequency sideband generated by the EOM was used as the signal pulse, which was also synchronized with the writing pulse. The frequency of the signal pulse was locked to the relevant WGM resonance of the silica microsphere with a Pound-Drever-Hall technique[8].

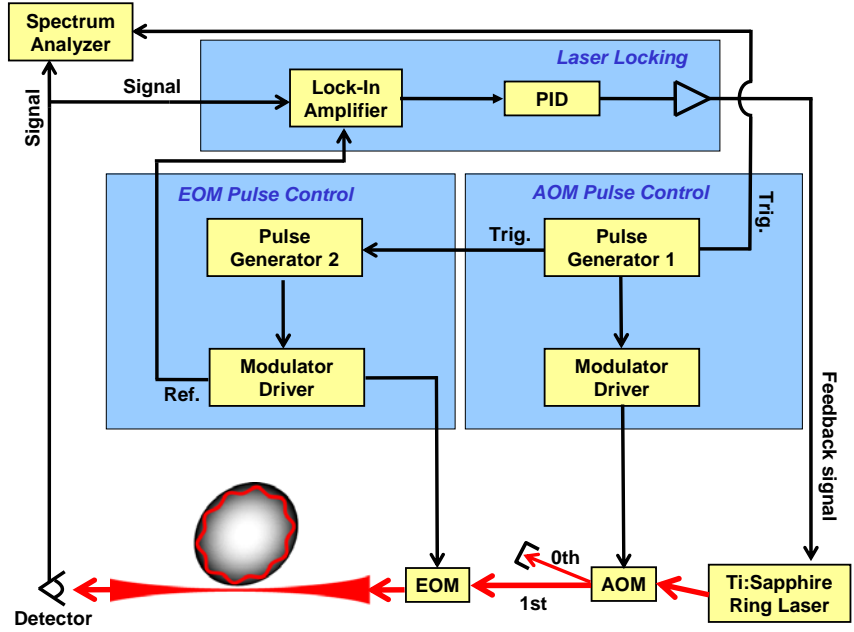

FIG. 3 Experimental setup, including diagrams for using the AOM and EOM to generate the writing, readout, and signal pulses, for locking the signal frequency to the WGM resonance, and for triggering the time-gated detection of the spectrum analyzer.

## 4. Heterodyne detection of the signal and retrieved pulses emitted from the resonator

We used direct heterodyne detection to measure the signal and retrieved pulses emitted from the silica optical resonator, as shown schematically in Fig. 4. For the heterodyne detection, the writing and readout pulses served as the local oscillator for the emitted signal and retrieved pulses, respectively.

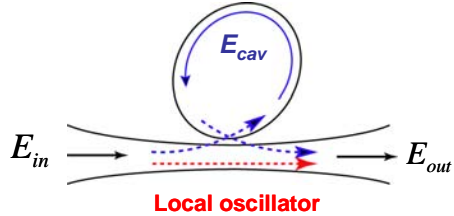

FIG. 4 Schematic of the direct heterodyne detection used for measuring signal and retrieved pulses emitted from the silica resonator.

We first discuss the heterodyne detection for the signal pulse. The well-known input-output relation that relates the intra-cavity field, $A_c$, to the incident and outgoing fields, $s_{in}$ and $s_{out}$ is given by[9]

$$s_{out}(\omega) = -s_{in}(\omega) + \sqrt{\kappa_{ex}} A_c(\omega) \qquad (5)$$

where the amplitudes are normalized such that $|s|^2$ is the power and $|A_c|^2$ is energy. The total incident field, which is phase-modulated by the EOM, can be written as

$$s_{in}(t) = s_w e^{-i[\omega_l t + M\sin(\Omega t)]} \approx s_w(1 + \frac{M}{2} e^{-i\Omega t} - \frac{M}{2} e^{i\Omega t}) e^{-i\omega_l t} \qquad (6)$$

where $s_w$ is the amplitude of the writing pulse, $M$ is the modulation depth, and $\Omega$ (which is near $\omega_m$) is the modulation frequency. For simplicity, we consider only the two primary sidebands. The higher frequency sideband at $\omega_l + \Omega$ is nearly resonant with the cavity mode and serves as the incident signal field. Note that the phase modulation alone does not give rise to a heterodyne-detected signal near frequency $\Omega$ since $|s_{in}(t)|^2 = s_w^2$. The power of the total outgoing field is given by

$$|s_{out}|^2 \approx -2 s_w \sqrt{\kappa_{ex}} \{\text{Re}[A_c(\omega_l + \Omega)]\cos(\Omega t) + \text{Im}[A_c(\omega_l + \Omega)]\sin(\Omega t)\}$$
$$+ (\text{DC term}) + (A_c^2 \text{ terms}) \qquad (7)$$

Note that we have ignored terms related to $A_c(\omega_l - \Omega)$, which (in the resolved-sideband limit) is far-off the optical resonance. The heterodyne-detected signal near frequency $\Omega$ thus measures directly the two quadratures of the signal pulse emitted from the optical cavity.

For the heterodyne detection of the retrieved pulse, the readout pulse, which serves as the local oscillator, is not phase-modulated by the EOM. The retrieved pulse is generated through an optomechanical process. The heterodyne detected signal is determined by an equation, which is the same as Eq. 7 except that $s_w$ is now replaced by the readout pulse amplitude, $s_r$.

We note that there is a thermal contribution in the heterodyne detection due to the Brownian motion of the mechanical oscillator. This thermal background is determined by the temperature of the mechanical oscillator. Under our experimental conditions, the thermal contribution is negligible, because of the relatively large signal and retrieval fields involved in the experiments.

## 5. Time-gated detection with a spectrum analyzer

We used a spectrum analyzer to measure the total powers of both quadratures of the signal and retrieved pulses. For time-resolved measurements, we operated the spectrum analyzer in a time-gated detection mode, where we measured the power spectrum as a function of the gate delay. The data acquisition occurs only during the detection gate. Figure 5 shows the timing for the detection process. For the measurement shown in Fig. 2a of the main manuscript, the time (x-axis) is taken to be delay between the rising edges of the writing/signal pulse and the detection gate. We took the spectrally-integrated spectral power density obtained from the power spectrum as the total power of the heterodyne signal. The time resolution of our measurements is limited by the resolution bandwidth (1 MHz) as well as by the gate length (3 μs). As shown in Fig. 2c of the main manuscript, the time-resolution of the gated-detection is approximately 3 μs, which is adequate for clearly separating the signal from the retrieved pulse.

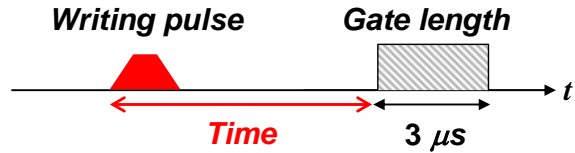

FIG. 5 Schematic for the gated detection. The power spectrum of the heterodyne signal is measured during the detection gate and as a function of delay between the rising edges of the writing pulse and the detection gate.

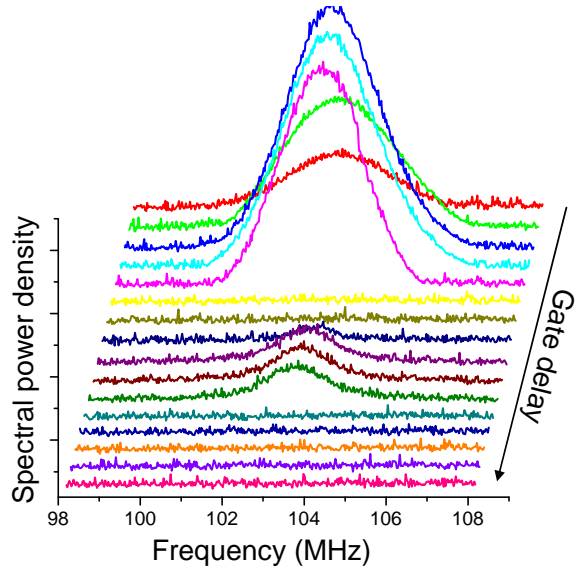

FIG. 6 Power spectra of the heterodyne-detected signal and retrieved pulses emitted from a silica microsphere. The gate delay between successive spectra is 1 μs. The timing for the writing, readout, and signal is the same as that in Fig. 2a in the main manuscript, except that the duration of the readout pulse is 3 μs. The spectrally-integrated spectral power density obtained from the power spectrum is the total power of the heterodyne signal.

Figure 6 shows, as an example, the power spectra obtained as a function of time for a light storage and retrieval measurement carried out in a silica microsphere with sample parameters and experimental conditions similar to those used in Fig. 2a of the main manuscript. The spectral linewidth of the resonance in these power spectra is in part due to the resolution band width of the gated-detection.